\documentclass[12pt]{article}

\begin{document}

\begin{center}
{\bf On massless and spinless particles with varying speed } \\
\vspace{5mm} S. I. Kruglov
\footnote{E-mail: krouglov@utsc.utoronto.ca}\\
\vspace{3mm}
\textit{Department of Chemical and Physical Sciences,\\ University of Toronto,
3359 Mississauga Rd. North,\\ Mississauga, Ontario, Canada L5L 1C6} \\
\vspace{5mm}
\end{center}

\begin{abstract}
The wave equation for spin-0 massless particles with the Lorentz violating term leading to varying speed of particles is considered. This equation is represented as the first-order 6$\times$6 matrix equation.
Solutions of the equation in the form of projection matrix are obtained. The Schr\"{o}dinger form of the equation is obtained and the quantum-mechanical Hamiltonian is found.
\end{abstract}

\section{Introduction}

The goal of this letter is to modify and investigate the wave equation for the simplest case of massless and spinless particles taking into account quantum gravity corrections. The space-time foam, within the Liouville string approach to quantum gravity \cite{Ellis}, influences on the propagation of low-energy particles. It should be mentioned that the string theory is the consistent quantum theory which can incorporate gravity. As a result of quantum gravity effect, within a non-critical formulation of string theory, space-time becomes a 'medium' (foam) and propagating quantum-mechanical particles interact with non-propagating gravitational stringy degrees of freedom. This leads to the modification of the dispersion relation for massless particles \cite{Ellis1} $E=|\textbf{k}|+\eta \textbf{k}^2/M_P$ ($M_P$ is the Plank mass, $\eta$ is a dimensionless parameter). The deformation of the dispersion relation is a consequence of the quantum friction of open dynamical system. As a result, due to interaction with foamy 'medium', massless particles propagate with a velocity $v=1+\eta E/M_P$.
which depends on energy. This phenomenon is connected with breakdown of conformal invariance in a quantum theory of non-critical strings, and to recover conformal symmetry we have to dress up the theory.  It should be mentioned, however, that the two-dimensional non-critical string \cite{Ellis} is a particular example of a stochastic formulation of quantum gravity. It was shown in \cite{Ellis} that in local scattering experiments the only propagating degrees of freedom, associated with observables, are massless scalar fields coupling space-time background (black holes). As a result, there are observable consequences of this interaction of foam and propagating fields.

The spontaneous violation of Lorentz and CPT symmetries within the string theory was also considered in \cite{Samuel}. The effective field theory with taking into consideration the Lorentz invariance violation (LIV) was investigated by \cite{Colladay} and bounds on LIV coefficients were obtained \cite{Russel}. Some models of LIV were studied in the photon sector \cite{Carroll} and fermion sector \cite{Ferreira}. The LIV models modify dispersion relations. Special form of the dispersion relation due to quantum gravity corrections was proposed in \cite{Amelino}, \cite{Smolin}:
\begin{equation}
p_0^2=\textbf{p}^2+m^2-\left(Lp_0\right)^\alpha\textbf{p}^2,
\label{1}
\end{equation}
where $p_0\equiv E$ is an energy and $\textbf{p}$ is a momentum of a
particle, the speed of light in vacuum $c=1$, and $L$ is the LIV parameter which is of the order of the Plank length $L_P=M_P^{-1}$. At $L>0$ particles propagate with the subluminal speed. The space-time foam Liouville-string models \cite{Ellis1} lead to the modified dispersion relation (1) with the parameter $\alpha=1$, and $m=0$. The wave equations for massive spinless particles realizing the dispersion equation (1) were considered in \cite{Ellis2}, \cite{Kruglov2}, and for massive spin-1/2 particles - in \cite{Kruglov3}.

The Euclidean metric is used here, we put $\hbar =c=1$, and Greek letters run 1,2,3,4 and Latin letters run 1,2,3.

\section{Wave equation and solutions}

I postulate the wave equation for massless and spinless particles as follows:
\begin{equation}
\left(\partial_{\mu}^2-iL\partial_i^2\partial_t\right)\Phi(x)=0,
\label{2}
\end{equation}
where $\partial_\mu=\partial/\partial x_\mu=\left(\partial/\partial x_i,\partial/(i\partial t)\right)$, $t$ is a time and $x_4=ix_0=it$. We treat Eq.(2) as an effective wave equation taking into consideration the LIV due to quantum gravity corrections and introducing preferred frame effects. We do not identify the field $\Phi(x)$ with a particular degree of freedom. The equation for massive spinless particles were considered in \cite{Ellis2}, \cite{Kruglov2}. The modified dispersion relation (1) with $m=0$ and $\alpha=1$ follows from Eq.(2) if one uses the plane-wave solution for positive energy $\Phi(x)=\Phi_0 \exp[i(\textbf{p}\textbf{x}-p_0x_0)]$. It should be noted that Eq.(2) is invariant under the rotation group but the invariance under the boost transformations is broken. The second term in Eq.(2) contains the LIV parameter $L$.
Eq.(2) is third order in derivatives and can be represented in the first-order form \cite{Kruglov1} and it follows from the system of first order equations
\[
\partial _\mu \Psi _\mu +\partial_4\widetilde{\Phi}=0,
\]
\begin{equation}
\partial_\mu\Phi+\kappa\Psi _\mu=0,
\label{3}
\end{equation}
\[
L\partial _m \Psi _m -\widetilde{\Phi}=0,
\]
where the parameter $\kappa$ with the dimension of ``mass" is introduced. The fields $\Psi_\mu$, $\Phi$, $\widetilde{\Phi}$ have the same dimension due to the mass parameter $\kappa$. Physical values should not depend on the $\kappa$. Let us introduce the wave function
\begin{equation}
\Psi (x)=\left\{ \Psi _A(x)\right\} =\left(
\begin{array}{c}
\Phi(x)\\
\Psi_\mu (x)\\
\widetilde{\Phi}(x)
\end{array}
\right),
\label{4}
\end{equation}
with index $A=(0,\mu ,\widetilde{0})$, $\Psi_0=\Phi$, $\Psi_{\widetilde{0}}=\widetilde{\Phi}$.
Exploring the elements of the entire matrix algebra $\varepsilon
^{A,B}$, with matrix elements and products \cite{Kruglov1}:
\begin{equation}
\left( \varepsilon ^{M,N}\right) _{AB}=\delta _{MA}\delta _{NB},
\hspace{0.5in}\varepsilon ^{M,A}\varepsilon ^{B,N}=\delta
_{AB}\varepsilon ^{M,N},
\label{5}
\end{equation}
where $A,B,M,N=(0,\mu ,\widetilde{0})$, Eqs.(3) can be
written as
\[
\biggl[\partial _\mu \left(\varepsilon ^{\mu,0}+ \varepsilon
^{0,\mu}+\delta_{\mu 4}\varepsilon ^{0,\widetilde{0}}-\kappa L\delta_{\mu m}\varepsilon ^{\widetilde{0},m}\right)
\]
\vspace{-8mm}
\begin{equation}
\label{6}
\end{equation}
\vspace{-8mm}
\[
+ \kappa\left(\varepsilon ^{\mu,\mu}+
\varepsilon ^{\widetilde{0},\widetilde{0}}\right)\biggr] _{AB}\Psi
_B(x)=0 .
\]
were the summation over all repeated indices is implied. Introducing the
$6\times 6$ matrices
\begin{equation}
\beta_\mu=\varepsilon ^{\mu,0}+ \varepsilon
^{0,\mu}+\delta_{\mu 4}\varepsilon ^{0,\widetilde{0}}-\kappa L\delta_{\mu m}\varepsilon ^{\widetilde{0},m},~~~
P=\varepsilon ^{\mu,\mu}+
\varepsilon ^{\widetilde{0},\widetilde{0}},
\label{7}
\end{equation}
with
\begin{equation}
\beta_m=\varepsilon ^{m,0}+ \varepsilon
^{0,m}-\kappa L\varepsilon ^{\widetilde{0},m},~~~
\beta_4=\varepsilon ^{4,0}+ \varepsilon
^{0,4}+\varepsilon ^{0,\widetilde{0}},
\label{8}
\end{equation}
and $P$ is a projection matrix, $P^2=P$, Eq.(6) takes the form of the first-order wave equation
\begin{equation}
\left( \beta _\mu \partial _\mu + \kappa P\right)
\Psi(x)=0 . \label{9}
\end{equation}
The 5-dimensional matrices $\beta_\mu^{(0)}=\varepsilon ^{\mu,0}+ \varepsilon
^{0,\mu}$ obey the Duffin$-$Kemmer$-$Petiau algebra and
enter the Lorentz covariant wave equation for scalar particles \cite{Kruglov1}.
The expression $J_{mn}=\varepsilon^{m,n}-\varepsilon^{n,m}$ represents the rotation group generators \cite{Kruglov1} and obeys the commutation relations
as follows:
\begin{equation}
\left[\beta_4,J_{m n}\right]=0,~~~\left[\beta_k,J_{m n}\right]=\delta_{k m}\beta_n
-\delta_{k n}\beta_m,~~~ \left[P,J_{m n}\right]=0.
\label{10}
\end{equation}
Thus, Eq.(9) is covariant under the  rotation group but not under the boost transformations. The matrices $\beta_\mu$ obey the algebra
\[
\beta_\mu\left(\beta_\nu\beta_\lambda\beta_\sigma+\beta_\sigma\beta_\lambda\beta_\nu\right)+
\beta_\lambda\left(\beta_\nu\beta_\mu\beta_\sigma+\beta_\sigma\beta_\mu\beta_\nu\right)+
\beta_\nu\left(\beta_\lambda\beta_\sigma\beta_\mu+\beta_\mu\beta_\sigma\beta_\lambda\right)
\]
\[
+\beta_\sigma\left(\beta_\lambda\beta_\nu\beta_\mu+\beta_\mu\beta_\nu\beta_\lambda\right)=
\delta_{\mu\nu}\left(\beta_\lambda\beta_\sigma+\beta_\sigma\beta_\lambda\right)+
\delta_{\lambda\nu}\left(\beta_\mu\beta_\sigma+\beta_\sigma\beta_\mu\right)
\]
\vspace{-8mm}
\begin{equation}
\label{11}
\end{equation}
\vspace{-8mm}
\[
+\delta_{\mu\sigma}\left(\beta_\lambda\beta_\nu+\beta_\nu\beta_\lambda\right)
+\delta_{\sigma\lambda}\left(\beta_\mu\beta_\nu+\beta_\nu\beta_\mu\right)
\]
\[
-\kappa L\biggl[\left(\delta_{\sigma m}\delta_{\nu 4}+\delta_{\nu m}\delta_{\sigma 4}\right)\left(\delta_{m\lambda}\beta_\mu +\delta_{m\mu}\beta_\lambda\right)
+ \left(\delta_{\lambda m}\delta_{\mu 4}+\delta_{\mu m}\delta_{\lambda 4}\right)\left(\delta_{m\sigma}\beta_\nu+\delta_{m\nu}\beta_\sigma\right)\biggr].
\]
which generalizes the Duffin$-$Kemmer$-$Petiau algebra.

The plane wave solution for the positive energies reads $\Psi(x)\sim \exp[i(\textbf{p}, \textbf{x}-p_0x_0)]$ and we have from Eq.(9)
\begin{equation}
\left( i\widehat{p} + \kappa P\right) \Psi (p)=0 , \label{12}
\end{equation}
where $\widehat{p}=\beta _\mu p _\mu$. From Eq.(11) one obtains the matrix equation for $\widehat{p}$
\begin{equation}
\widehat{p}^4 -p^2\widehat{p}^2+\kappa Lp_4\textbf{p}^2\widehat{p}=0 ,
\label{13}
\end{equation}
with notations: $p^2=\textbf{p}^2-p_0^2$, $p_4=ip_0$.
One can verify with the help of (13) that the matrix
\begin{equation}
\Lambda = i\widehat{p} + \kappa P
\label{14}
\end{equation}
satisfies the equation as follows:
\begin{equation}
\Lambda\left(\Lambda-\kappa\right)\left[\Lambda\left(\Lambda-\kappa\right)^2+p^2\left(\Lambda-\kappa\right)-
i\kappa Lp_4\textbf{p}^2 \right]=0.
\label{15}
\end{equation}
From Eq.(15), we obtain (off-shell) the solutions to Eq.(12) in the form
\begin{equation}
\Pi=N\left(\Lambda-\kappa\right)\left[\Lambda\left(\Lambda-\kappa\right)^2+p^2\left(\Lambda-\kappa\right)-
i\kappa Lp_4\textbf{p}^2 \right]
 \label{16}
\end{equation}
and $\Lambda\Pi=0$; $N$ is the normalization constant. Thus, each column of the matrix $\Pi$ is the solution to Eq.(12). Requirement that $\Pi$ is the projection matrix \cite{Fedorov}, $\Pi^2=\Pi$, gives
the normalization constant
\begin{equation}
N=\frac{1}{\kappa^2\left(p^2-Lp_0\textbf{p}^2\right)} .
\label{17}
\end{equation}
It should be noted that on-shell, when Eq.(1) is satisfied with $m=0$, $\alpha=1$, the minimal matrix equation (15) becomes
\begin{equation}
\Lambda^2\left(\Lambda-\kappa\right)\left[\left(\Lambda-\kappa\right)^2+p^2 \right]=0.
\label{18}
\end{equation}
In this case zero eigenvalues of the matrix $\Lambda$ are degenerated and it is impossible to obtain solutions to Eq.(12) in the form of projection operators \cite{Fedorov}. Note that when Eq.(1) is satisfied ($m=0$, $\alpha=1$), the denominator of (17) becomes zero.

\section{Schr\"{o}dinger form}

Now we consider the Schr\"{o}dinger form of Eq.(9). One can rewrite Eq.(9) as follows:
\begin{equation}
i\beta _4\partial _t\Psi (x)=\biggl (\beta
_m\partial_m+\kappa P\biggr )\Psi (x).
 \label{19}
\end{equation}
It is easy to verify that the matrix $\beta_4$ obeys the matrix equation
\begin{equation}
 \beta _4^4=\beta_4^2 ,
\label{20}
\end{equation}
so that the matrix $\Sigma=\beta_4^2$ is the projection operator, $\Sigma^2=\Sigma$.
The dynamical components of the wave function $\Psi(x)$ read $\phi
(x)=\Sigma \Psi(x)$. We define the projection operator
\begin{equation}
\Omega =1-\Sigma = \varepsilon ^{\widetilde{0},\widetilde{0}}+\varepsilon ^{m,m}-\varepsilon ^{4,\widetilde{0}},
\label{21}
\end{equation}
which obeys the equality $\Omega^2=\Omega$. Then non-dynamical components of the wave function $\Psi (x)$ are $\chi(x)=\Omega\Psi (x)$. Multiplying Eq.(19) by the matrix $\beta_4$, and taking into account the equation $\beta_4\beta_m\beta_4=0$, one finds
\begin{equation}
i\partial _t \phi(x)=\beta_4\beta
_m\partial_m\chi(x)+\kappa\beta_4P\Psi(x).
\label{22}
\end{equation}
The relation $\Psi(x)=\phi(x)+\chi(x)$ is valid because $\Sigma+\Omega=1$.
After multiplying Eq.(19) by the matrix $\Omega$, and taking into account the equations $\Omega \beta_4
=0$, $\beta_4P\Omega=0$, we obtain
\begin{equation}
\Omega \beta_n \partial_n\Psi(x) +\kappa\chi(x)=0 .
\label{23}
\end{equation}
Expressing the component $\chi$ from Eq.(23)
\begin{equation}
\chi(x)=-\frac{1}{\kappa}\Omega \beta_n \partial_n\Psi(x)
\label{24}
\end{equation}
and putting it into Eq.(22), with the aid of the equations $\beta_4\beta_m\Omega\beta_n\Omega=0$, $\beta_4P\Omega=0$, one finds the Schr\"{o}dinger equation
\begin{equation}
i\partial _t\phi (x)=\biggl (
-\frac{1}{\kappa}\beta_4\beta_m\partial_m\Omega\beta_n \partial_n+\kappa\beta_4P\biggr)\phi(x).
\label{25}
\end{equation}
The wave function $\phi$ has only two non-zero components
\begin{equation}
\phi (x)=\left(
\begin{array}{c}
\Phi(x)\\
\Psi_4(x)+\widetilde{\Phi}(x)
\end{array}
\right),
\label{26}
\end{equation}
corresponding to states with positive and negative energies.
We can represent Eq.(25) in the form
\begin{equation}
i\partial _t\phi (x)=\mathcal{H}\phi(x),
\label{27}
\end{equation}
where the Hamiltonian is
\begin{equation}
\mathcal{H}=\frac{1}{\kappa}\left(\kappa L\varepsilon^{0,0}-\varepsilon^{4,0}
\right)\partial_m^2+\kappa\left(\varepsilon^{0,4}+\varepsilon^{0,\widetilde{0}}\right).
\label{28}
\end{equation}
With the help of Eq.(5), the Hamiltonian (28) can be written as
\begin{equation}
\mathcal{H}=\left(
\begin{array}{cc}
L\partial_m^2&\kappa \\
-(1/\kappa)\partial_m^2&0
\end{array}\right).
\label{29}
\end{equation}
Eq.(27), using Eqs.(26),(29), can be written as a system of equations
\[
i\partial _t\Phi (x)= \kappa\left(\Psi_4(x)+\widetilde{\Phi}(x)\right)+L\partial_m^2\Phi(x) ,
\]
\vspace{-7mm}
\begin{equation} \label{30}
\end{equation}
\vspace{-7mm}
\[
i\partial _t\left(\Psi_4(x)+\widetilde{\Phi}(x)\right)=
-\frac{1}{\kappa}\partial_m^2\Phi(x) .
\]
Eqs.(30) may be found from Eqs.(3),
replacing components $\Psi_m (x)=-(1/\kappa)\partial_m \Phi(x)$. One can notice that Eqs.(27), (30) include only components with the time derivatives. The Hamiltonian form Eq.(27) has an advances compared to Eq.(9) because it is $2\times 2$ matrix-differential equation and the wave function has only two dynamical components connected with positive and negative ehergies. The Hamiltonian (29) in the momentum space $\partial_\mu\rightarrow ip_\mu$ satisfies the equation as follows:
\begin{equation}
\mathcal{H}^2+L\textbf{p}^2\mathcal{H}-\textbf{p}^2I_2=0,
\label{31}
\end{equation}
with $I_2$ being $2\times 2$ identity matrix. It follows from Eq.(31) that the eigenvalue of the
Hamiltonian $p_0$ obeys the dispersion equation (1) at $m=0$, $\alpha=1$.  It should be noted that the
Schr\"{o}dinger form is convenient for solving problems with interacting particles. One can introduce in the Hamiltonian the potentials connecting particle interaction with external fields.

\section{Conclusion}

In this letter we consider an effective theory of massless and spinless particles which takes into consideration quantum-gravitational friction and leads to LIV. The wave equation proposed allow us to investigate quantum processes with taking into account quantum gravity corrections.
The spinless and massless particles of the proposed wave equation possess varying speed depending on the energy. This is a consequence of interaction of particles with foamy 'medium'. Such effective interaction is due to the evolution of the quantum-mechanical system during the time which is longer as compared with the Planck time. As a result, this leads to open quantum-mechanical system and the interaction with environment can be treated as an interaction with space-time foam. This approach corresponds to a stochastic formulation of quantum gravity.

For low energy the particle speed is the light speed $c$, but for high energy the speed is less than $c$ for $L>0$. The first-order wave equation formulated is convenient for different applications. The solutions in the form of projection matrix (16) can be used for calculations of quantum processes with massless and spinless particles. The Schr\"{o}dinger form of the equation obtained can be used for quantum-mechanical calculations. The Hamiltonian obtained allows us to introduce the interaction of light particles with external fields in a simple manner as it is a $2\times 2$ matrix. The first-order wave equation and solutions obtained as well as the Hamiltonian found are the basis for the investigation of different quantum processes of interacting particles taking into account quantum gravity corrections. When spin effects of massless particles (for example, photons, neutrinos) are negligible the proposed wave equation can be used for studying the propagation and interaction with different particles. At present energies the LIV effects are suppressed by the Planck scale $M_P=1.22\times 10^{19}$ GeV and there are not signs yet of LIV in experiments. The analysis of bounds on the LIV parameter $L$ from experimental data and consequences of a proposed model will be further investigated.


\begin{thebibliography}{99}

\bibitem{Ellis} J. R. Ellis, N. E. Mavromatos and D. V. Nanopoulos, Phys. Lett. \textbf{B293},  37 (1992)
    [arXiv:hep-th/9207103];
 Chaos, Solitons and Fractals \textbf{10}, 345 (1999) [arXiv:hep-th/9805120].

\bibitem{Ellis1} G. Amelino-Camelia, J. R. Ellis, N. E. Mavromatos and D. V. Nanopoulos, 
Int. J. Mod. Phys. \textbf{A12}, 607 (1997) [arXiv:hep-th/9605211].

\bibitem{Samuel} V. A. Kostelecky and S. Samuel, Phys. Rev. \textbf{D39}, 683 (1989); Phys. Rev. Lett.  \textbf{66}, 1811 (1991); V. A. Kostelecky and R. Potting, Nucl. Phys. \textbf{B359}, 545 (1991); Phys. Rev. \textbf{D51}, 3923 (1995) [hep-ph/9501341 [hep-ph]].

\bibitem{Colladay} D. Colladay and V. A. Kostelecky, Phys. Rev. \textbf{D58}, 116002 (1998) [hep-ph/9809521];
S. R. Coleman and S. L. Glashow, Phys. Rev. \textbf{D59}, 116008 (1999) [hep-ph/9812418].

\bibitem{Russel} V. A. Kostelecky and N. Russell, Rev. Mod. Phys. \textbf{83}, 11 (2011)
[arXiv:0801.0287 [hep-ph]].

\bibitem{Carroll} S. M. Carroll, G. B. Field and R. Jackiw, Phys. Rev. \textbf{D41}, 1231 (1990);
R. Jackiw and V. A. Kostelecky, Phys. Rev. Lett. \textbf{82}, 3572 (1999) [hep-ph/9901358 [hep-ph]];
S. I. Kruglov, Phys. Lett. \textbf{B652}, 146 (2007) [arXiv:0705.0133 [hep-ph]]; V. A. Kostelecky and M. Mewes, Phys. Rev. \textbf{D80}, 015020 (2009) [arXiv:0905.0031 [hep-ph]].

\bibitem{Ferreira} M. Chaichian, W. F. Chen and R. Gonzalez Felipe, Phys. Lett. \textbf{B503}, 215 (2001) [hep-th/0010129 [hep-th]];
 M. M. Ferreira, Jr. and F. M. O. Moucherek, Int. J. Mod. Phys. \textbf{A21},
6211 (2006) [hep-th/0601018 [hep-th]]; P. A. Bolokhov and M. Pospelov, Phys. Rev. \textbf{D77}, 025022 (2008) [hep-ph/0703291 [hep-ph]]; H. Belich et al, Eur. Phys. J. \textbf{C62}, 425 (2009) [arXiv:0806.1253 [hep-th]]; A. Kostelecky and M. Mewes, Phys. Rev. \textbf{D85}, 096005 (2012) [arXiv:1112.6395 [hep-ph]].

\bibitem{Amelino} G. Amelino-Camelia, Int. J. Mod. Phys. \textbf{D11}, 35 (2002) [arXiv:gr-qc/0012051];
 New J. Phys. \textbf{6}, 188 (2004) [arXiv:gr-qc/0212002].

\bibitem{Smolin} J. Magueijo and L. Smolin, Phys. Rev. Lett., \textbf{88}, 190403 (2002) [arXiv:hep-th/0112090];
Phys. Rev. \textbf{D67}, 044017 (2003) [arXiv:gr-qc/0207085].

\bibitem{Ellis2} J. Ellis, N. E. Mavromatos and A. S. Sakharov, Astropart. Phys. \textbf{20}, 669 (2004)
[arXiv:astro-ph/0308403].

\bibitem{Kruglov2} S. I. Kruglov, Mod. Phys. Lett. \textbf{A28}, No. 6, 1350014 (2013) [arXiv:1207.6573 [hep-th]].

\bibitem{Kruglov3} S. I. Kruglov,  Phys. Lett. \textbf{B718}, 228 (2012) [arXiv:1210.0509 [gr-qc]].

\bibitem{Kruglov1} S. I. Kruglov, Symmetry and Electromagnetic Interaction of Fields with
Multi-Spin (Nova Science Publishers, Huntington, New York, (2001)).

\bibitem{Fedorov} F. I. Fedorov,  Sov. Phys. - JETP \textbf{35}(8) (1959), 339 (Zh.
Eksp. Teor. Fiz. \textbf{35} (1958), 493).


\end{thebibliography}
\end{document}